\documentclass[aps,prl,reprint,superscriptaddress, longbibliography]{revtex4-2}

\usepackage{amsmath}
\usepackage{amssymb}
\usepackage{amsfonts}
\usepackage{units}
\usepackage{color}
\usepackage[utf8]{inputenc}
\usepackage[T1]{fontenc}
\usepackage{psfrag,graphics,graphicx,subfigure,hyperref}
\usepackage[usenames,dvipsnames]{xcolor}
\usepackage{amsbsy}
\usepackage{ulem}
\usepackage{float}
\newcommand{\Kcbar}{\overline{K}_\mathrm{c}}

\newcommand{\Etot}{ {\cal E}}

\newcommand{\MV}[1]{\textcolor{black}{#1}} 
\newcommand{\lolo}[1]{\textcolor{black}{#1}} 
\begin{document}
\graphicspath{{Figs/}}

\title{Beyond linear stability: Heterogeneity-induced fingering of crack fronts}



\author{Manish Vasoya}
\email[]{manishvasoya36@gmail.com}
\affiliation{Department of Material Science and Engineering, Rutgers University, Piscataway, NJ 08854, USA}
\author{Laurent Ponson}
\email[]{laurent.ponson@sorbonne-universite.fr}
\affiliation{Institut Jean Le Rond d'Alembert (UMR 7190), CNRS and Sorbonne Universit\'e, 75005 Paris, France}

\author{Veronique Lazarus}
\email[]{veronique.lazarus@ensta.fr}
\affiliation{IMSIA, ENSTA,  CNRS,  EDF,   Institut Polytechnique de Paris, 91120 Palaiseau, France}


\date{\today}

\begin{abstract}

We investigate the stability of elastic interfaces beyond the linear regime by considering penny-shaped crack fronts propagating through toughness heterogeneities. Using fracture mechanics simulations, we drive crack fronts through arrays of obstacles with tunable toughness contrast. At low contrast, the crack front stiffness remains finite and stabilizes front perturbations. Above a critical threshold, however, the stiffness vanishes and the front destabilizes into long fingers that evolve into daughter cracks propagating between obstacles while the original crack remains pinned. Near threshold, the crack front response displays the characteristic square-root scaling behavior of classical saddle-node bifurcations. Yet, our analysis reveals a fundamentally different mechanism: the stable energy-minimizing crack-front configuration disappears without colliding with an unstable counterpart. Instead, the instability originates from a global loss of Griffith-compatible equilibria governed by the nonlocal interactions along the crack front. Beyond fracture mechanics, these findings point toward a broader class of collective global bifurcations in nonlocal elastic interfaces and may help rationalize the brittle-to-quasibrittle transition in heterogeneous solids.

\end{abstract}


\maketitle

\begin{figure}[h!]
\begin{center}
\includegraphics[width=1\linewidth]{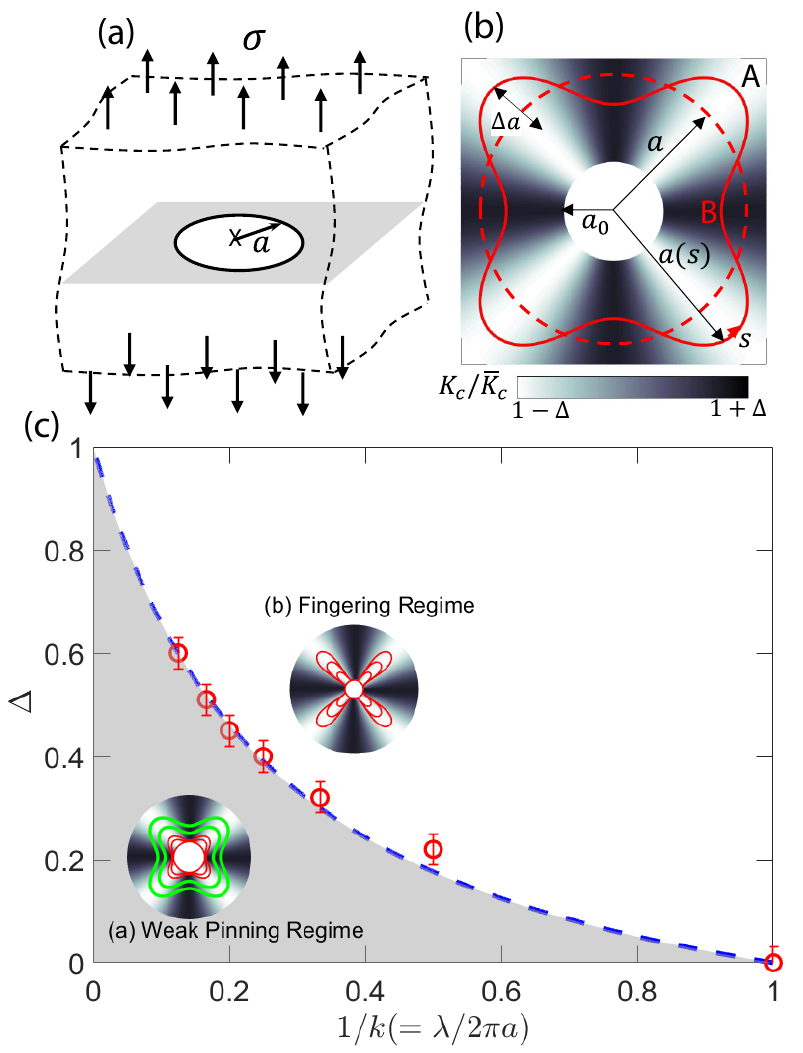}
\caption{\footnotesize Fingering instability of penny-shaped cracks fronts: (a) Three-dimensional configuration of a penny-shaped crack propagating in an infinite solid under a remotely applied stress $\sigma$. (b) Periodic toughness landscape used to deform the crack front, characterized by toughness contrast $\Delta$ and wavelength $2\pi a/k$, where $a$ is the mean crack radius and $k$ is the number of obstacles (here $k=4$). (c) Phase-diagram in the ($\Delta$, $1/k$) plane showing the weak pinning (gray) and fingering (white) regime. Open circles indicate the fingering threshold measured from fracture mechanics simulations, while the dashed blue line shows the fit $\Delta_\mathrm{c} (1/k) \simeq (1-1/k)/(1+3.65/k)$.}
\label{fig:summary}
\end{center}
\end{figure}
Driven interfaces are ubiquitous and underlie phenomena as diverse as wetting~\cite{Joanny}, magnetism~\cite{Zapperi2}, solid phase transformation~\cite{Dondl}, growth processes in biology~\cite{Mobius}, multiphase flow~\cite{Saffman,ganesh}, dislocation loops~\cite{Gao91,VatChi22,ZhangSills}, earthquakes \cite{GaoLeeRic91,FinSve14, CohFinKol15, GviKamAddFin25}, and adhesion \cite{Sanner}. 
In all these physical situations, the geometrical stability of the interface, i.e., its ability to keep an overall straight configuration despite the destabilizing effects of material heterogeneities, loading conditions, thermal fluctuations, etc., is a central issue, as the interface behavior at small scales drives the transport or mechanical properties at large scales. For small amplitude perturbations, a linear stability analysis is generally sufficient to infer the interface stability and to some extent, characterize the unstable modes. For example, this approach accounts for the viscous Saffman-Taylor fingering between two immiscible fluids~\cite{Saffman} or the destabilization of a peeling front in a confined adhesive layer~\cite{Saintyves,Biggins,Ghatak,Ghatak3}. However, this method gives no hint on the interface response to {\it finite-amplitude} perturbations that may result from naturally strong heterogeneities or obstacles, as for instance those designed to achieve new properties~\cite{Xia,Hossain}.

Here, we investigate this problem in the context of fracture, using crack fronts propagating in heterogeneous brittle solids~\cite{Lawn,Alava2,Bonamy6} as a paradigmatic realization of elastic interfaces. To probe their nonlinear response, we consider penny-shaped cracks drove through periodic arrays of obstacles with tunable size and toughness. While this configuration was previously used to explore the relationship between local and global fracture properties~\cite{Vasoya4}, we focus here on the {\it stability} of the crack front.

At low toughness contrasts, we recover the results of the linear stability analysis that predicts stable crack growth~\cite{Rice4, Gao3}. Beyond a critical toughness contrast, however, the crack front destabilizes into fingers that grow unstably. Near threshold, the crack-front dynamics displays the hallmark scaling signatures of a classical saddle-node bifurcation. Surprisingly, however, our analysis provides no evidence for the unstable branch required by the conventional fold scenario. Instead, the instability appears to result from the disappearance of Griffith-compatible energy-minimizing crack-front configurations. Finally, we discuss how this mechanism may extend to a broad class of elastic interfaces and how it could offer a route for rationalizing the brittle-to-quasibrittle transition in heterogeneous materials.

Penny-shaped cracks propagating in brittle solids provide a prototypical realization of elastic interfaces when dissipative failure processes remain confined to the vicinity of the crack tip~\cite{Rice3,Barenblatt}. This looped interface behaves then like an elastic line with long-range elasticity that emerges from the material bulk elasticity~\cite{Rice4, Schmittbuhl4}. In the reference configuration with homogeneous material properties and remote uniform tensile loading (Fig. \ref{fig:summary}a), the crack front adopts a circular shape of radius $a$ and the driving force acting on it is uniform. Its magnitude is set by the Stress Intensity Factor (SIF), $K_\circ = 2 \, \sigma \, \sqrt{a/\pi} $ where $\sigma$ is the tensile stress applied remotely, or equivalently, by the rate $G_\circ = - \delta \mathcal{E}_\mathrm{m}/\delta \mathcal{S} = K_\circ^2\,(1-\nu^2)/E$ of mechanical energy released for a uniform crack advance $\delta a = \delta \mathcal{S}/2\pi a$, where $\mathcal{S}$ is the crack surface area, and $E$ and $\nu$ are the Young's modulus and the Poisson's ratio, respectively~\cite{Irwin}. The crack evolution is then inferred from the comparison of $K_\circ$ with the material toughness $K_\mathrm{c}$, or equivalently, of $G_\circ$ with the fracture energy $G_\mathrm{c} = \delta \mathcal{E}_\mathrm{diss}/\delta \mathcal{S} = K_\mathrm{c}^2\,(1-\nu^2)/E$ where $\mathcal{E}_\mathrm{diss}$ is the energy dissipated by fracture through the so-called Griffith criterion $\dot{a} \, (G_\circ-G_c) = 0$ that derives from thermodynamics of irreversible processes~\cite{Rice5}. This criterion implies that during crack propagation, $\dot{a} > 0$, the mechanical energy released during crack growth is dissipated into fracture energy, a condition that writes as $G_\circ = G_\mathrm{c}$~\cite{Griffith,Rice5}. Note that we assume here quasi-static crack propagation and neglect any rate-dependence of the fracture energy. Rewritten in terms of SIF, the propagation criterion writes as $\dot{a} \, (K_\circ-K_\mathrm{c}) = 0$. The concepts of driving force and growth resistance, defined here within a thermodynamic framework, can be generalized to any interface problem.

Under the effect of material heterogeneities or heterogeneous loading conditions, the interface geometry becomes more complex and the driving force $K(s)$ varies with the position $s$ along the interface (\MV{Fig. \ref{fig:summary}b}). The Griffith criterion then applies locally along the front~\cite{Wang, Laz23MealorII2023}. For cracks with small perturbations $\delta a(s) = a(s) - a$, $\left|\delta a(s)\right| \ll a$, \lolo{where $a=\sqrt{\mathcal{S}/\pi}$,} the corresponding variations of the SIF $\delta K(s) = K(s) - K_\circ$ can be determined using a perturbation-based formula $\delta K(s) = \mathcal{W} \ast \delta a(s) $ that involves a long-range kernel $\mathcal{W} \sim 1/s^2$~(see Refs.~\cite{Lazarus2, LazJTCAM} for  reviews).

To explore the linear stability of moving interfaces in general, and crack fronts in particular,  small harmonic perturbations  $\delta a^\mathrm{(k)}(\theta) = -\Delta a/2 \, \cos(k \theta)$ of amplitude $\Delta a \ll a$ and wavelength $\lambda = 2\pi a/k$ where $k \geq 2$ is the wave number are considered. Applying the linearized formula that relates $\delta K(\theta)$ to $\delta a(\theta)$ using the expression of the kernel $\mathcal{W}$  for penny-shaped cracks~\cite{Gao3} provides the driving force perturbations $\delta K(\theta)/K_\circ =  (k-1) \, \delta a^\mathrm{(k)}(\theta)/2a$  that also follow sinusoidal variations, but with an opposite sign. As a result, the most advanced regions of the front located in $\theta = 0, \, 2\pi/k, ... 2\pi(k-1)/k$ are also those with the smallest driving force, ensuring that any small front perturbations, that can be decomposed as a sum of harmonic modes $\delta a^\mathrm{(k)}(\theta)$, vanish, accounting for the circular shape adopted by penny-shaped cracks growing in homogeneous materials~\cite{Gao3}. 

To explore whether penny-shaped cracks remain stable in the presence of perturbations of {\it finite} amplitude, we must go beyond linear stability analysis. We propose herein to trigger large perturbations using a heterogeneous toughness field $K_\mathrm{c}(\theta) = \Kcbar \left[1 + \Delta \cos(k\theta)\right]$, that consists in $k$ obstacles of toughness contrast $\Delta \in [0 \; 1]$ (Fig.~\ref{fig:summary}b), $\Kcbar$ being the mean toughness \cite{Vasoya4}. Starting from a circular configuration of radius $a_0$, the crack grows and deforms in a flower shape with $k-$petals. The size of the crack can be described by its surface area $\mathcal{S}$ or equivalently by the radius $a=\sqrt{\mathcal{S}/\pi}$ of the circle of same surface \cite{DavLaz22}, and its deformation by the petal amplitude $\Delta a$, as schematized in Fig.~\ref{fig:summary}b. To ensure quasi-static propagation, the applied stress $\sigma$ is adjusted at each time step so that at least one point along the front (here the points located in the weak regions $\theta = \pi/k,$ $3\pi/k$, ...  $(2k-1)\pi/ k$ of the fracture plane) verifies $K = K_\mathrm{c}$ while $K \leq K_\mathrm{c}$ elsewhere. To carry this non-linear stability analysis and compute $K(s)$ along the front, one must resort to either higher-order analytical theory~\cite{Lebihain} or numerical methods~\cite{BowOrt90,Lazarus3}. These require the determination of the perturbations of the kernel $ \mathcal{W}$ that is obtained by an analog formula to $\delta K(s) = \mathcal{W} \ast \delta a(s)$~\cite{Rice7}. In practice, linear approximations of $\delta K(s)$ and  $\delta \mathcal{W}$ are added incrementally during the deformation of the front, starting from the initial circular configuration~\cite{PlaneCracks, Lazarus3, Vasoya3}.

For weak obstacles $\Delta \ll 1$,  the method is equivalent to a linear stability analysis. Indeed, the stationary modes of growth satisfying $K(\theta) = K_\mathrm{c}(\theta)$ correspond to the harmonic perturbations $\delta a^\mathrm{(k)}(\theta)$ introduced previously, with $ \Delta a /a =  4 \Delta/(k-1) $. However, considering tougher obstacles, the method allows to study the stability of penny-shaped cracks to perturbations of any amplitude. The results of our numerical investigation, comprising hundreds of simulations, are compiled in the phase diagram of Fig.~\ref{fig:summary}c in the normalized obstacle width $1/k = \lambda/2\pi a$ versus toughness contrast $\Delta$ plane. For small toughness contrasts, $\Delta < \Delta_c(1/k)$, the crack front adopts a stationary shape represented in green in the inset (a) of Fig. \ref{fig:summary}c. In this so-called {\it weak pinning} regime, the propagation threshold $K(s) = K_\mathrm{c}(s)$ is reached everywhere along the front, after an initial transient phase during which the crack advances in the weakest zones only (successive red front positions). Moreover, the effective toughness that defines the load required to propagate the crack through the relation $K^{\mathrm{eff}}_{\mathrm{c}} \equiv  2\sigma \sqrt{a/\pi}$ turns out to be \lolo{nearly equal to} the mean toughness $K^{\mathrm{eff}}_{\mathrm{c}} \simeq \Kcbar$ \cite{Vasoya4}.
Conversely, for large toughness contrasts, $\Delta > \Delta_c(1/k)$, a {\it fingering} regime sets in: Some regions of the front never reach the propagation threshold and thus remain trapped by the obstacles while the rest satisfies locally to the propagation criterion $K = K_\mathrm{c}$ and thus keeps propagating yielding infinitely growing fingers invading the weakest regions of the fracture plane, as shown in inset(b) of Fig. \ref{fig:summary}c. An interesting limit deserves attention: As ${k \to \infty}$, the critical toughness contrast goes to one. This corresponds to infinitely tough obstacles, highlighting the stability of semi-infinite cracks, even to perturbations of finite amplitude, as taking the limit $k \to \infty \Leftrightarrow \lambda \ll a$ amounts to considering a straight crack front. This is consistent with the observation that straight crack fronts stiffen as they deform beyond the linear regime~\cite{Leblond2, Vasoya2}.

\begin{figure}[h]
\begin{center}
\includegraphics[width=1\linewidth]{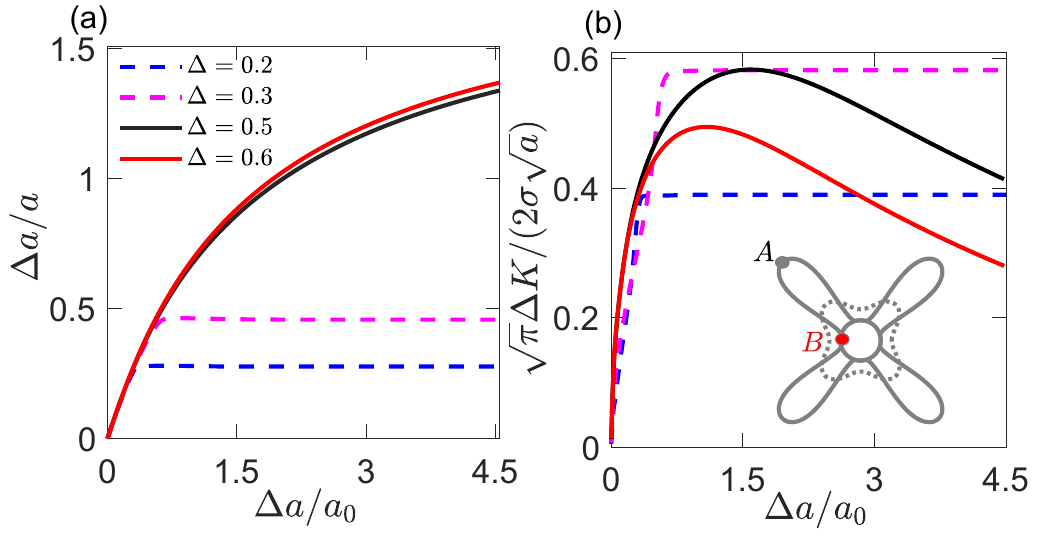}
\caption{\footnotesize 
Local interpretation of the fingering instability from the driving-force variations along the crack front, shown here for $k=4$ obstacles with critical toughness contrast $\Delta_\mathrm{c} \simeq 0.40$.
(a) Evolution of the normalized petal length during crack propagation. For $\Delta < \Delta_\mathrm{c}$, it saturates to a stationary value $(\Delta a/a)_\mathrm{s}$, whereas it grows continuously in the fingering regime for $\Delta > \Delta_\mathrm{c}$. (b) Evolution of the normalized SIF contrast between the pinned regions $B$ and the regions $A$ propagating between obstacles.
Inset: typical crack front geometries in the weak pinning (dashed line) and fingering (solid line) regimes.}
\label{fig:FIC}
\end{center}
\end{figure}

Both crack growth regimes can be differentiated kinematically from the evolution of the dimensionless petal length $\Delta a/a$ shown in Fig.~\ref{fig:FIC}a. In the weak pinning regime, $\Delta a/a$ saturates to a stationary value $(\Delta a/a)_\mathrm{s}$, while the normalized petal length grows continuously in the fingering regime. The underlying mechanism is studied in Fig.~\ref{fig:FIC}b that shows the difference of normalized SIF $\Delta K/\overline{K}_{\mathrm{c}} \equiv \left[K(B)-K(A)\right]/\overline{K}_{\mathrm{c}}$ between the regions $B$ of the front trapped by the obstacles and the regions $A$ growing in-between. As the initial crack configuration is circular, the SIF is initially uniform, so $\Delta K = 0$. However, as the front propagates in $A$ in the weakest regions of the fracture plane, the SIF increases in $B$ owing to the front elasticity. In the weak pinning regime, this results in an increase of $\Delta K$ until it reaches a plateau, $\Delta K/\overline{K}_{\mathrm{c}} = 2\Delta$, reminiscent of the stationary state.

Unstable cracks display a strikingly different behavior, as $\Delta K$ first increases and then starts decreasing before the untrapping threshold $\Delta K/\overline{K}_{\mathrm{c}}=2\Delta$ can be reached. This unexpected drop may result from the shielding effect produced by the large petals, which behave like individual cracks on their own (see the shape of the crack shown with {\it solid} line in inset of Fig. \ref{fig:FIC}b). It explains why regions $B$ lag behind the rest of the crack and ultimately remain trapped by the obstacles as an array of nearly independent cracks (or petals) continues to grow. As a support to this interpretation, one notices that $\Delta K$ starts decreasing when $\Delta a \simeq a_0$, {\it i.e.} when daughter cracks reach a size similar to their mother crack. This suggests that the long-range elastic restoring forces that maintain penny-shaped cracks in an overall circular shape vanish for perturbations {\it larger} than the crack itself.

To get deeper insights into the nature of this instability, we now recast the formation of fingers in terms of energy landscape and variational approach to fracture~\cite{FraMar98}.
We focus on the total energy $\Etot$ defined as the sum of the elastic energy and the fracture cost. For a circular crack in a homogeneous medium, it reads $\mathcal{E}_\circ=\pi a^2 \overline{G}_{\rm c} - 8/3 \pi \, \sigma^2 a^3 (1-\nu^2)/E$~\cite{sup}.
It reaches a maximum at $a_\circ(\sigma)=\pi/4 (\overline{K}_{\rm c}/\sigma)^2$ corresponding to the crack size satisfying the Griffith condition $G_\circ = G_\mathrm{c}$. We now introduce front perturbations and focus on the energy variations $\Delta \mathcal{E}_\circ = \mathcal{E}-\mathcal{E}_\circ$. We seek for the front shape that minimizes $\Delta \mathcal{E}_\circ$ at constant crack surface area under a constant applied stress $\sigma$~\cite{sup}.

For small perturbations $\delta a(s)$, the variation of elastic energy is $\int_{\mathcal{F}} G(s) \delta a(s) ds$ while the variation of fracture cost is $\int_{\mathcal{F}} G_c(s) \delta a(s) ds$, where both integrals are taken along the unperturbed crack front $\mathcal{F}$. Hence, the energy variation $\Delta\mathcal{E}_\circ$ can be obtained incrementally using the same perturbation method used previously to update $K$. 
In the weak pinning regime, we force petals to grow {\it beyond} their stationary amplitude $(\Delta a/a)_\mathrm{s}$ by using a toughness contrast $\Delta_\mathrm{driving}  > \Delta_\mathrm{c}$ once the front has reached its stationary shape while keeping the primary toughness field to evaluate the variations of fracture energy. The energy landscape obtained by this protocol turns out to be qualitatively independent of $\Delta_\mathrm{driving}$~\cite{sup}.

\begin{figure}[h!]
\centerline{\includegraphics[width=\linewidth]{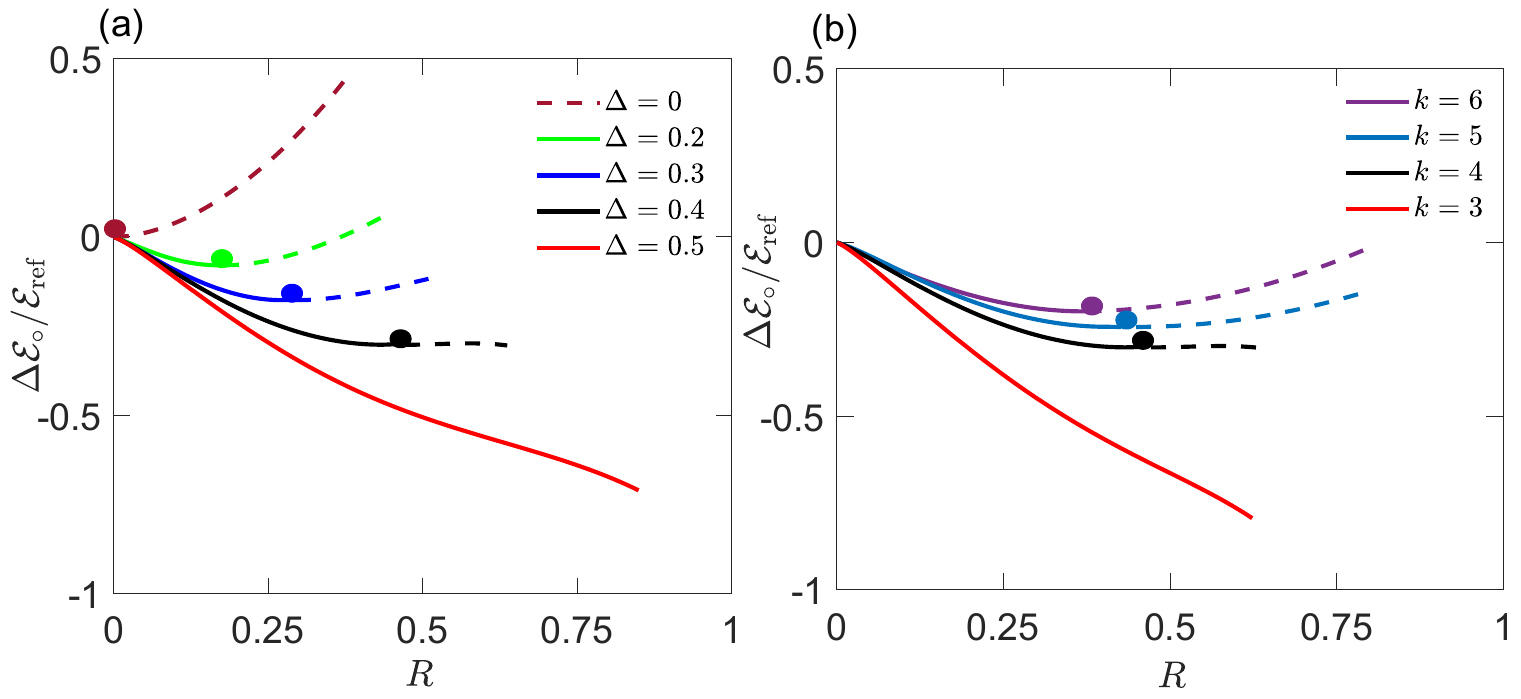}}
\caption{\footnotesize Global interpretation of the fingering instability from the variations of the total energy with normalized petal size $\Delta a/a$. (a) Energy landscapes for $k=4$ ($\Delta_{c} \simeq 0.40$) for different toughness contrasts $\Delta$. (b) Energy landscapes for $\Delta = 0.4$ ($k_\mathrm{c} \simeq 4$) for different obstacle numbers $k$. Solid lines are obtained using the prescribed toughness contrast $\Delta$ while dashed lines are obtained using $\Delta_{\rm driving} = 0.6 > \Delta_c$. In the weak pinning regime, the energy exhibits a minimum corresponding to stationary crack configurations, whereas it decreases monotonically in the fingering regime. 
}
\label{fig:energy}
\end{figure}

The variations $\Delta \Etot_\circ$ of total energy normalized by the reference energy $\mathcal{E}_{\mathrm{ref}}=\overline{G}_{\mathrm{c}} \mathcal{S}$ is shown in Fig.~\ref{fig:energy} as a function of the petal aspect ratio $R \equiv \Delta a/\lambda$ where $\lambda = 2\pi a/k$ is the petal width, for different toughness contrasts $\Delta$ and obstacle numbers $k$. Focusing first on the weak pinning regime $\Delta < \Delta_\mathrm{c} \Leftrightarrow k > k_\mathrm{c}$, we observe an initial decrease of the total energy for small petal size $\Delta a/a < (\Delta a/a)_\mathrm{s}$, followed by an increase of energy for larger petals. The energy-minimizing configurations marked by dots in Fig.~\ref{fig:energy} correspond to stationary crack fronts of amplitude $(\Delta a/a)_\mathrm{s}$ satisfying the Griffith criterion $G(s)=G_\mathrm{c}(s)$, highlighting the equivalence between a global energy minimization approach and Griffith's local propagation criterion~\cite{Lawn,FraMar98,Wang}.

Strikingly, as we approach the fingering threshold $\Delta \rightarrow \Delta_\mathrm{c} \Leftrightarrow k \rightarrow k_\mathrm{c}$, the potential well progressively flattens and eventually disappears. As shown in the supplemental materials~\cite{sup}, the relaxation time required for a crack front to retrieve its stationary shape after being perturbed diverges as $\tau \propto (\Delta_\mathrm{c} - \Delta)^{-1/2}$, revealing a critical slowing down commonly observed near saddle-node bifurcations.

\begin{figure}[h!]
\centerline{\includegraphics[width=1.1\linewidth]{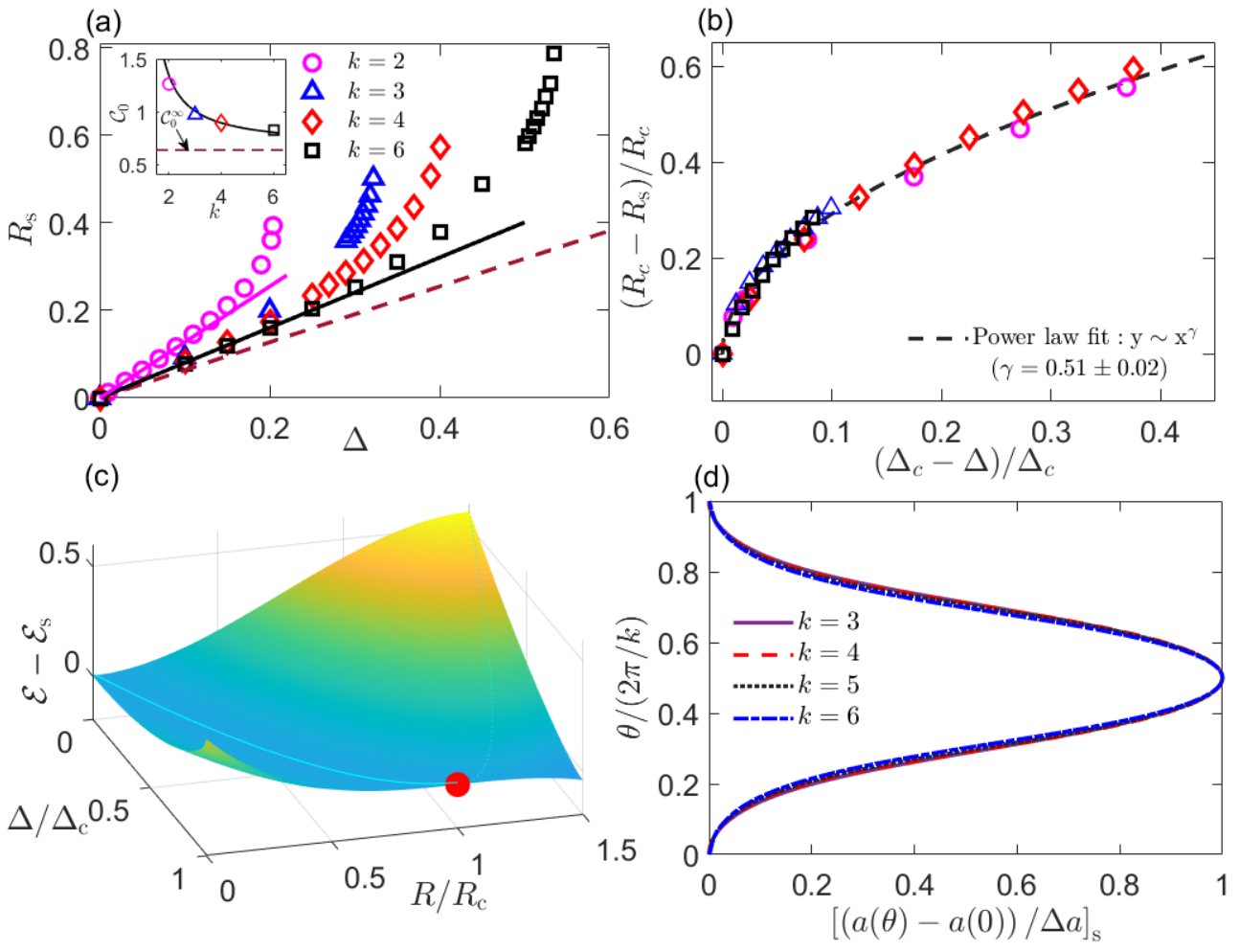}}
\caption{\footnotesize Universal behavior of penny-shaped cracks near the fingering threshold: (a) Variations of the stationary petal aspect ratio $R_\mathrm{s} \equiv \Delta a_\mathrm{s}/\lambda =  k (\Delta a/a)_\mathrm{s} / 2\pi$ with toughness contrast $\Delta$ for different obstacle numbers $k$. Inset: comparison between the linear crack front compliance $\mathcal{C}_0(k) \equiv {\rm d}R/{\rm d}\Delta|_{\Delta \to 0}$ and linear theory predictions. (b) Collapse of normalized stationary petal aspect ratio $(R_\mathrm{c} - R_\mathrm{s})/R_\mathrm{c}$ onto a single master curve when plotted as a function of the normalized distance $(\Delta_\mathrm{c}-\Delta)/\Delta_\mathrm{c}$ to the fingering threshold. A power-law fit (dashed line) yields $\gamma = 0.51 \pm 0.02$, consistent with square-root scaling. (c) Schematic energy landscape in the ($R$, $\Delta$) plane. The bifurcation point ($R_\mathrm{c}$, $\Delta_\mathrm{c}$) is indicated by the red dot, while the stable stationary branch is shown by the solid line. Although the landscape exhibits a maximum, it does not correspond to Griffith-compatible stationary crack-front configurations; (d) Universal critical finger shape at threshold for different obstacle numbers $k$.}
\label{fig:saddle}
\end{figure}

We now explore in Fig.~\ref{fig:saddle} how the stationary petal size, {\it i.e.} the order parameter of the bifurcation, varies with the toughness contrast $\Delta$, its control parameter. Panel~(a) shows the petal aspect ratio $R_{\mathrm{s}}$ of the stationary crack configurations as a function of the toughness contrast. For $\Delta \ll \Delta_\mathrm{c}$, the petal aspect ratio increases linearly, $R_{\mathrm{s}} = \mathcal{C}_0(k) \, \Delta$, where $\mathcal{C}_0(k)$ is the linear front compliance. The linear stability analysis predicts $\mathcal{C}_0(k)= 2 k/(\pi (k-1))$~\cite{Gao3}, in agreement with our simulations as shown in inset of Fig.~\ref{fig:saddle}a. Interestingly, the compliance $\mathcal{C}_0^\infty = 2/\pi$ of a semi-infinite crack~\cite{Gao,Vasoya} is retrieved when $k \rightarrow \infty$, confirming that the behavior of straight cracks is recovered in this limit.

With increasing obstacle toughness, the crack displays a non-linear response featured by an increasing compliance $\mathcal{C}(\Delta) = {\rm d}R_{\mathrm{s}}/{\rm d} \Delta > \mathcal{C}_0$, until it diverges for $R_{\mathrm{s}} \rightarrow R_\mathrm{c} \Leftrightarrow \Delta \rightarrow \Delta_\mathrm{c}$ at the onset of fingering. This non-linear response $R_\mathrm{s}(\Delta)$ is shown in Fig.~\ref{fig:saddle}b in the normalized form $(R_\mathrm{c}-R_{\mathrm{s}})/R_\mathrm{c}$ versus $(\Delta-\Delta_\mathrm{c})/\Delta_\mathrm{c}$ for different values of $k$. Remarkably, all data collapse on a single master curve, revealing a power-law behavior $\left(R_\mathrm{c}-R_{\mathrm{s}}\right)/R_\mathrm{c} \simeq \left[(\Delta_\mathrm{c}-\Delta)/\Delta_\mathrm{c}\right]^{\gamma}$ with exponent $\gamma = 0.51 \pm 0.02$. This implies that the crack front compliance diverges as $\mathcal{C} \propto 1/\sqrt{\Delta_c-\Delta}$ upon approaching the onset of fingering for $\Delta \to \Delta_c$, or equivalently, that the front stiffness vanishes as $\mathcal{K} \equiv 1/\mathcal{C} \propto \sqrt{\Delta_\mathrm{c} - \Delta}$.

To elucidate this behavior, we consider the total energy $\Etot(\Delta,R)$ introduced previously that is represented schematically in Fig.~\ref{fig:saddle}c in the (petal aspect ratio $R/R_c$, toughness contrast $\Delta/\Delta_c$) plane. Because all stationary crack-front configurations satisfy the Griffith criterion, moving from one nearby stationary configuration to another leaves the total energy unchanged. The stable branch therefore forms an effectively isoenergetic valley of energy-minimizing crack-front configurations. Incorporating this property into a Taylor expansion of the energy near the bifurcation point yields $\Etot(\Delta, R_\mathrm{s}) \simeq  \Etot(\Delta_\mathrm{c},R_\mathrm{c}) +  \,  \partial^2 \Etot/\partial \Delta \partial R \, (\Delta - \Delta_\mathrm{c}) (R_\mathrm{s}- R_\mathrm{c}) + 1/6 \, \partial^3 \Etot/\partial R^3 \, (R_\mathrm{s} - R_\mathrm{c})^3 + ...$ where only the leading-order terms have been retained and all derivatives are evaluated at the critical point~\cite{sup}. Balancing the two contributions immediately gives $R_\mathrm{c} - R_\mathrm{s} \propto \sqrt{\Delta_\mathrm{c} - \Delta}$ in agreement with the simulation results of Fig~\ref{fig:saddle}b.

At first sight, the fingering instability exhibits the hallmark features of a classical saddle-node bifurcation, including the square-root scaling of both the perturbation amplitude and the relaxation time~\cite{Kuznetsov}. Such behavior would normally be associated with the collision and disappearance of stable and unstable branches. Yet, several observations suggest a fundamentally different picture.

First, all stationary crack-front configurations satisfying the Griffith criterion remain isoenergetic. As a consequence, a hypothetical Griffith-compatible unstable branch should lie at the same energy level as the stable branch while appearing as an ensemble of maxima on the energy landscape of Fig.~\ref{fig:saddle}c, two features that do not seem compatible with each other. Second, throughout the stable regime, the observed branch acts as a unique attractor with no identifiable unstable counterpart, as even strongly fingered crack fronts ultimately relax toward the Griffith-compatible stationary configuration~\cite{sup}. Taken together, these observations suggest that fingering does not originate from a local stable–unstable branch annihilation, but rather from a global loss of Griffith-compatible equilibria. Thenon-local elasticity of the crack front plays a major role in this mechanism, as it imposes constraints on the {\it global} geometry of the stable configurations.

This suggests that nonlocal elastic interfaces may host a broader class of collective global instabilities, fundamentally distinct from the conventional local bifurcation paradigm of stable-unstable branch annihilation. Remarkably, near threshold, the crack front converges toward a universal critical shape shown in Fig.~\ref{fig:saddle}d ~\cite{sup}. This indicates that at the onset of fingering, the crack front loses memory of the details of the toughness map and becomes mainly governed by the nonlocal elasticity of the interface.

To conclude, we emphasize the distinctive and potentially generic nature of the fingering instability uncovered in crack fronts. By introducing obstacles with tunable toughness contrast, we drove penny-shaped crack fronts far beyond their linear response regime. Above a critical threshold, the front becomes unstable at a finite deformation and develops long fingers that eventually form arrays of daughter cracks. From a local perspective, this transition originates from the progressive vanishing of the crack-front stiffness, leaving the interface unable to suppress large front distortions. From a global perspective, however, the instability corresponds to the disappearance of the entire branch of stationary Griffith-compatible crack-front configurations, without any identifiable unstable counterpart.

Taken together, these observations suggest that nonlocal elastic interfaces may undergo instabilities fundamentally distinct from conventional local bifurcations. Yet, despite their different origin, these transitions retains the hallmark phenomenology of classical saddle-node bifurcations, including the square root scaling of both the perturbation amplitude and relaxation time near threshold.

An important extension of the present work would consist in relaxing the assumption of a connected crack front and allowing for crack nucleation. This could provide a route toward describing the brittle-to-quasibrittle transition within the broader context of collective global bifurcations in nonlocal elastic interfaces.

{\it Acknowledgments -} The authors thank Prof. Jean-Baptiste Leblond, Dr. Mathias Lebihain, Dr. Patrick Ribeiro, and Prof. K. Ravi-Chandar for fruitful discussions.

\bibliography{bibfracture}

\end{document}


\title{Supplemental material "Beyond linear stability: Heterogeneity-induced fingering of cracks"}


\author{Manish Vasoya}
\email[]{manishvasoya36@gmail.com}
\affiliation{Department of Material Science and Engineering, Rutgers University, Piscataway, NJ 08854, USA}
\author{Laurent Ponson}
\email[]{laurent.ponson@sorbonne-universite.fr}
\affiliation{Institut Jean Le Rond d'Alembert (UMR 7190), CNRS and Sorbonne Universit\'e, 75005 Paris, France}

\author{Veronique Lazarus}
\email[]{veronique.lazarus@ensta.fr}
\affiliation{IMSIA, ENSTA,  CNRS,  EDF,   Institut Polytechnique de Paris, 91120 Palaiseau, France}

\date{\today}
\maketitle
\appendix 

\section{Energy landscape calculation}

The aim of this appendix is to present the method used to compute the evolution of the total energy with the perturbation size as presented in Fig. 3 of the main text. In a nutshell, the method already used in \cite{Vasoya4} is employed to compute the evolution of the energy release rate and of the crack surface area. It is  based on the incremental Bueckner\cite{Bue87}-Rice\cite{Rice7} small perturbation approach \cite{Lazarus3} that provides the distribution of SIF along a perturbed crack front. In practice, this method is implemented using the Planecracks code~\cite{PlaneCracks}.

Our primary objective is to compute the variations  of total energy during the evolution of a penny-shaped crack of unit surface area under constant fixed load. The variations in total energy at each incremental step read as, 
 \begin{equation}\label{eq:appVarE0}
 \Delta \mathcal{E} = \displaystyle \sum_{0}^{\Delta a} \int_{\mathcal{F}} \left[G_c(s)-G(s)\right] \delta a(s) ds,
\end{equation}
where $G_c(s)-G(s)$ denotes the decrease in elastic energy and the increase in the cost of fractures as the crack is perturbed by $\delta a(s)$.  
 
\begin{figure}[h]
\centerline{\includegraphics[width=1\linewidth]{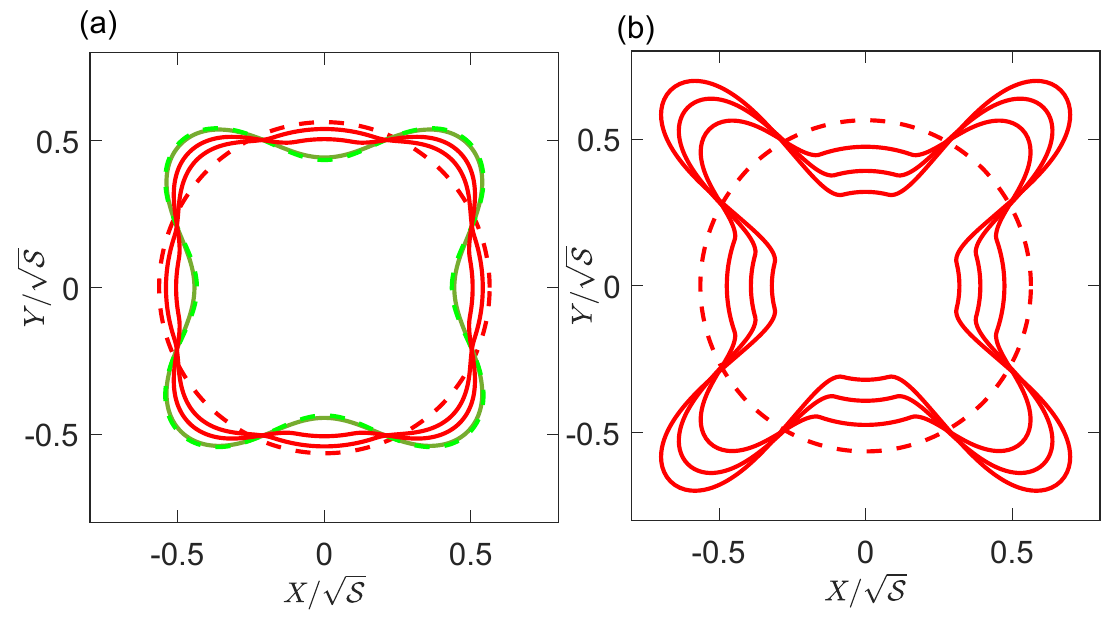}}
\caption{Successive normalized crack fronts with the mean crack radius imposed at $1/\sqrt{\pi}$ (dashed red circle): (a) Successive normalized crack fronts in the weak pinning regime with $k=4, \ {\rm and} \ \Delta=0.3$. The two solid red fronts correspond to the transient regime before reaching the stationary Griffith-compatible crack front geometry shape in green. (b) Non-stationary normalized crack front evolution in the fingering regime for $k=4, \ {\rm and} \ \Delta=0.5$.}
\label{fig:app1}
\end{figure}

Within the heterogeneous toughness map considered here, the crack front propagates resulting in an increase of the total crack surface area. To ensure a constant crack surface to focus on the sole effect of the shape of the front, we normalize the radius $a$ of the propagating crack by the square root of the crack surface area $\sqrt{\mathcal{S}}$,  so that the corresponding surface area is equal to unity throughout crack evolution. 
The Figure~\ref{fig:app1} represents the successive normalized crack fronts shown in insets of Fig. 1(c) of the main text. This highlights that the crack surface area is kept constant while the crack front deforms.

To compute the energy variations for a unit cracked surface area, all length-scales are normalized by $\sqrt{\mathcal{S}}$, yielding
\begin{equation}\label{eq:appVarE1}
\frac{\Delta \mathcal{E}}{\mathcal{S}} = \displaystyle \sum_{0}^{\Delta a/\sqrt{\mathcal{S}}} \int_{\mathcal{F}} \left[G_c(s)-G(s)\right] \frac{\delta a(s)}{\sqrt{\mathcal{S}}} \frac{ds}{\sqrt{\mathcal{S}}}.
\end{equation}
To render the previous equation fully non-dimensional, we further divide both sides by the mean fracture energy $\overline{G}_c$, yielding
\begin{equation}\label{eq:appVarE}
\frac{\Delta \mathcal{E}}{\mathcal{E}_{\rm ref}} = \displaystyle \sum_{0}^{\Delta a/\sqrt{\mathcal{S}}} 
\int_{\mathcal{F}} \left[\frac{G_c(s)}{\overline{G}_c}-\frac{G(s)}{\overline{G}_c}\right] \frac{\delta a(s)}{\sqrt{\mathcal{S}}} \frac{ds}{\sqrt{\mathcal{S}}},
\end{equation}
where, $\mathcal{E}_{\rm ref}=\overline{G}_c\mathcal{S}$ is the reference energy defined in the main text. The first term, $G_c(s)/\overline{G}_c$, can be expressed as a function of $K_c(s) = \overline{K}_c [1+\Delta \cos (k\theta(s))]$~\cite{Vasoya3}. Similarly, one can replace the second term of Eq.~\ref{eq:appVarE} by 
\begin{equation}\label{eq:G}
\frac{G(s)}{\overline{G}_c} = \frac{1}{\mathcal{A}_{\Delta}} \frac{K^2(s)}{\overline{K}_c^2},
\end{equation}
where $\mathcal{A}_{\Delta}=1+\Delta^2/2$. 

Finally, we impose that the effective SIF, $K_{\rm eff},$ corresponding to an equivalent circular crack of radius $a$ remains constant during crack propagation and equal to mean toughness $\overline{K}_c$,
\begin{equation} \label{eq:fixedSIF}
2\sigma \sqrt{a/\pi}=2\sigma \sqrt{\frac{\sqrt{\mathcal{S}}}{\pi\sqrt{\pi}}}=\overline{K}_c.
\end{equation}
Owing to the relation $K_\mathrm{eff} \simeq \bar{K}_\mathrm{c}$ shown in Ref.~\cite{Vasoya4} in the weak pinning regime and the unit crack area imposed previously, this amounts to maintain the applied load $\sigma$ constant. Inserting this last expression  in Eq.~\eqref{eq:G} yields
\begin{equation}\label{eq:normG}
\frac{G(s)}{\overline{G}_c} = \frac{1}{\mathcal{A}_{\Delta}} \frac{K^2(s)}{\overline{K}_c^2}
= \frac{1}{\mathcal{A}_{\Delta}} \frac{\pi\sqrt{\pi}}{4}\frac{\widehat{K}^2(s)}{ \sqrt{\mathcal{S}}}.
\end{equation}
where $\widehat{K}=\frac{K(s)}{\sigma}$ is the SIF corresponding to a unit applied stress. Its evolution during the deformation of the crack front is computed using PlaneCracks~\cite{PlaneCracks}.\\

\begin{figure}[h]
\centerline{\includegraphics[width=.8\linewidth]{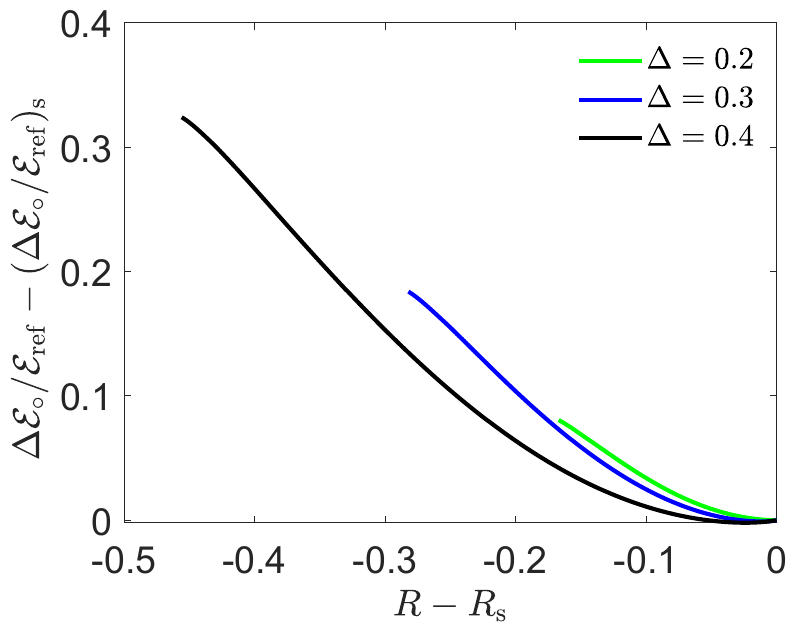}}
\caption{Energy landscape taking the stationary front configuration as the reference energy level (replot of Fig.~3a of main text).}
\label{fig:app2}
\end{figure}

As the crack front starts deviating from its initial circular configuration, during the transient regime, the total energy variation $\Delta \mathcal{E}/\mathcal{E}_{\rm ref}$ decreases. When the crack front reaches its stationary Griffith-compatible configuration, the bracket term in Eq.~\ref{eq:appVarE} goes to zero, and $\Delta \mathcal{E}/\mathcal{E}_{\rm ref}$ reaches a minimum. The total energy becomes independent of the toughness contrast, as the stationary crack fronts form an isoenergetic valley of energy-minimizing configurations. This property is illustrated in Fig.~\ref{fig:app2}, where the energy-landscapes of Fig.~3a of the main text are replotted after subtracting the reference energy level of the stationary configuration.
We observe that, initially, the curves corresponding to different contrast values have distinct origins, and as we approach the stationary configuration, they all asymptotically converge to the same minimum energy state. 
\vspace*{-0.3cm}
\section{Influence of the choice of the driving contrast value used to deform the crack front beyond its stationary configuration}
This appendix demonstrates that the choice of the driving contrast value, $\Delta_{\rm driving}$, used to deform the crack front beyond its stationary or equilibrium configuration, only induces minor quantitative changes, while leaving the qualitative behavior — and hence the conclusions of the paper — unchanged.

\begin{figure}[h]
\centerline{\includegraphics[width=.8\linewidth]{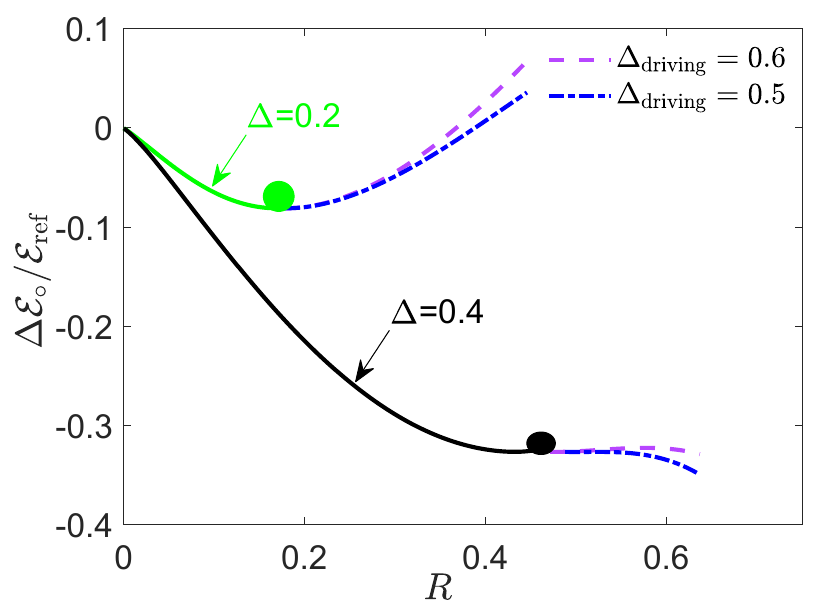}}
\caption{Influence of the choice of the larger toughness contrast value  $\Delta_{\rm driving}$  used to deform a given stationary crack front shape above the minimum equilibrium point.}
\label{fig:app3}
\end{figure}

We consider two cases of the toughness field as representative: $\Delta = 0.2$ and $0.4$ with $k=4$. As discussed, after the crack front reaches its stationary configuration, $\Delta a/a$ stops evolving further, and hence, $\Delta \mathcal{E}/\mathcal{E}_{\rm ref}$ ({\it green} and {\it black} solid curves for $\Delta=0.2$ and $0.4$, respectively, ending with filled circles in Fig. \ref{fig:app3}). To further deform the stationary crack configuration, we use a larger contrast $\Delta_{\rm driving} > \Delta_c$. To calculate the energy landscape along this artificially deformed crack front, we use the contrast value $\Delta$ as in the regular front deformation. 

This raises the question: How does $\Delta_{\rm driving}$ affect the energy landscape of Fig. 3? As shown in the previous section, regardless of the toughness contrast value, the energy landscape evolves into a minimum-energy state with the same asymptote as the crack geometry approaches its stationary configuration. Similarly, irrespective of the $\Delta_{\rm driving}$ value, the energy landscape evolves from the minimum-energy state, with the same asymptote, as the crack front deforms beyond its stationary configuration. As representative example (Fig. \ref{fig:app3}), we have used two values of $\Delta_{\rm driving} = 0.5$, $0.6$ ({\it dashed} magneta and {\it dashed-dot} blue curves, respectively) for each $\Delta=0.2$, $0.4$ case. The energy landscape in both cases looks similar and most importantly, it evolves from the minimum energy state with the same asymptote as we deform the crack front from its stationary configuration. This confirms that the stationary crack configuration corresponds to a minimum energy state and that the main conclusions of the letter remains independent of the choice of $\Delta_{\rm driving}$.

\vspace*{-0.9cm}
\section{Expression of the total energy for an unperturbed penny-shaped crack}
The reference energy $\mathcal{E}_0$ used in the energy landscapes of the main manuscript is obtained from the fracture cost $\int_0^a G_\mathrm{c} 2\pi a \, da = \pi a^2 G_\mathrm{c} $ and the mechanical energy $\mathcal{E}_\mathrm{m} = \int_0^a G(a) 2\pi a \, da $. Using the energy release rate of an unperturbed penny-shaped crack, $G_0 = (1-\nu^2) K_0^2/E = (1-\nu^2) 4 \sigma^2a/\pi E$, gives $\mathcal{E}_\mathrm{el} 
  = - 8/3 \pi \, \sigma^2 a^3 (1-\nu^2)/E$. The total energy therefore reads $\mathcal{E}_0 = \pi a^2 G_\mathrm{c} -  8/3 \pi \, \sigma^2 a^3 (1-\nu^2)/E$.

\vspace*{0.9cm}
\section{Scaling behavior near the fingering threshold}

\subsection{Stationary crack-front deformation}
We derive the scaling law $R_\mathrm{c} - R_\mathrm{s} \propto \sqrt{\Delta_\mathrm{c}-\Delta}$ observed numerically in Fig.~4(b) of the main text. To this end, we fix the number $k$ of obstacles, and expand the energy landscape $\mathcal{E}(\Delta,R_s)$ shown in Fig.~4(c) about the bifurcation point $(\Delta_\mathrm{c},R_\mathrm{c})$
\begin{equation}
\label{eq:energy1}
\begin{aligned}
\mathcal{E}(\Delta,R_\mathrm{s}) & = \mathcal{E}(\Delta_\mathrm{c},R_\mathrm{c}) + \frac{\partial \mathcal{E}}{\partial \Delta} (\Delta - \Delta_\mathrm{c}) + \frac{\partial \mathcal{E}}{\partial R} (R_\mathrm{s}-R_\mathrm{c}) \\
& + \frac12 \frac{\partial^2 \mathcal{E}}{\partial \Delta^2}(\Delta-\Delta_\mathrm{c})^2 +  \frac12 \frac{\partial^2 \mathcal{E}}{\partial R^2}(R_\mathrm{s}-R_\mathrm{c})^2 \\
& + \frac{\partial^2 \mathcal{E}}{\partial \Delta \partial R}(\Delta - \Delta_\mathrm{c})(R_\mathrm{s} - R_\mathrm{c})  \\
& + \frac16\frac{\partial^3 \mathcal{E}}{\partial \Delta^3}(\Delta-\Delta_\mathrm{c})^3 + \frac16 \frac{\partial^3 \mathcal{E}}{\partial R^3}(R_\mathrm{s} - R_\mathrm{c})^3  \\
& + \frac12 \frac{\partial^3 \mathcal{E}}{\partial \Delta^2 \partial R}(\Delta - \Delta_\mathrm{c})^2 (R_\mathrm{s}-R_\mathrm{c}) \\
&  + \frac12 \frac{\partial^3 \mathcal{E}}{\partial \Delta \partial R^2}(\Delta-\Delta_\mathrm{c}) (R_\mathrm{s}-R_\mathrm{c})^2 + ...
\end{aligned}
\end{equation}
where all derivatives are evaluated at $(\Delta_\mathrm{c},R_\mathrm{c})$. Stationary configurations minimize the energy with respect to $R$, as shown in Fig.~3 of the main text, yielding 
\begin{equation}
\left. \frac{\partial \mathcal{E}}{\partial R}\right |_{\Delta_\mathrm{s}, R_\mathrm{s}} = 0.
\end{equation}
These Griffith-compatible configurations are also isoenergetic along the stationary branch, yielding
\begin{equation}
\left. \frac{\partial \mathcal{E}}{\partial \Delta}\right |_{\Delta_\mathrm{s}, R_\mathrm{s}} = 0.
\end{equation}
Taking the limit $(\Delta,R_s) \to (\Delta_c,R_c)$ therefore gives
\begin{equation}
\left. \frac{\partial \mathcal{E}}{\partial \Delta} \right |_{\Delta_\mathrm{c}, R_\mathrm{c}} = \left. \frac{\partial \mathcal{E}}{\partial R}\right |_{\Delta_\mathrm{c}, R_\mathrm{c}} = 0.
\end{equation}

At the bifurcation point, the energy minimum disappears as the landscape loses its curvature in the $R$-direction (see Fig.~3 of the main text), yielding
\begin{equation}
\left. \frac{\partial^2 \mathcal{E}}{\partial R^2} \right |_{\Delta_\mathrm{c}, R_\mathrm{c}} = 0.
\end{equation}

Keeping the remaining terms in the energy expansion of Eq.~\eqref{eq:energy1} and using the isoenergetic property of the stationary branch $\mathcal{E}(\Delta,R_\mathrm{s}) = \mathcal{E}(\Delta_\mathrm{c},R_\mathrm{c})$ yields
\begin{equation}
\begin{aligned}
& \frac12 \frac{\partial^2 \mathcal{E}}{\partial \Delta^2}(\Delta-\Delta_\mathrm{c})^2 + \frac{\partial^2 \mathcal{E}}{\partial \Delta \partial R}(\Delta - \Delta_\mathrm{c})(R_\mathrm{s} - R_\mathrm{c})  \\
& + \frac16\frac{\partial^3 \mathcal{E}}{\partial \Delta^3}(\Delta-\Delta_\mathrm{c})^3 + \frac16 \frac{\partial^3 \mathcal{E}}{\partial R^3}(R_\mathrm{s} - R_\mathrm{c})^3 \\
& + \frac12 \frac{\partial^3 \mathcal{E}}{\partial \Delta^2 \partial R}(\Delta - \Delta_\mathrm{c})^2 (R_\mathrm{s}-R_\mathrm{c}) \\
&  + \frac12 \frac{\partial^3 \mathcal{E}}{\partial \Delta \partial R^2}(\Delta-\Delta_\mathrm{c}) (R_\mathrm{s}-R_\mathrm{c})^2 = 0
\label{eq_dvt}
\end{aligned}
\end{equation}
where all derivatives are evaluated at $(\Delta_\mathrm{c},R_\mathrm{c})$.

To identify the dominant balance, we assume $(R_\mathrm{c} -R_\mathrm{s}) \propto (\Delta_c -\Delta)^{\alpha}$ with $0 < \alpha < 1$. Substitution into Eq.~\eqref{eq_dvt} shows that the leading contributions are the mixed term $\frac12\frac{\partial^2 \mathcal{E}}{\partial \Delta \partial R} (\Delta - \Delta_\mathrm{c})(R_\mathrm{s} - R_\mathrm{c})$ scaling as $(\Delta-\Delta_c)^{1+\alpha}$ and the cubic term in $R$, scaling as $(\Delta-\Delta_c)^{3\alpha}$. Balancing these two powers gives $1+\alpha=3\alpha$, hence $\alpha = 1/2$. We therefore obtain
\begin{equation}
(R_\mathrm{c}-R_\mathrm{s}) \propto \sqrt{\Delta_\mathrm{c}-\Delta}
\end{equation}

Beyond characterizing the stationary branch, this square-root scaling has a direct consequence for the relaxation dynamics, as shown in the next section.

\subsection{Relaxation dynamics}
As a direct consequence of the square-root scaling of the stationary deformation amplitude, the relaxation time associated with the recovery of stationary Griffith-compatible crack-front configurations diverges as $\tau \propto (\Delta_\mathrm{c} - \Delta)^{-1/2}$ near the fingering threshold. We derive this result below.

We start from the linearized crack -front evolution equation
\begin{equation}
\frac{\partial a}{\partial t} \propto G[a(\theta,t)] - G_\mathrm{c}(\theta,a(\theta,t))
\end{equation}
which follows from the linearization of the kinetic law $G_\mathrm{c}(v)$ around vanishing crack speed~\cite{Chopin5}.

Using
\begin{equation}
G[a(\theta)] = - \frac{\delta \mathcal{E}_\mathrm{m}}{\delta a}
\end{equation}
and the fracture cost variation
\begin{equation}
\delta \mathcal{E}_\mathrm{f} = \int_\mathcal{F} G_\mathrm{c}(\theta,a(\theta)) \delta a \, d\theta,
\end{equation}
the functional derivative of the total energy satisfies
\begin{equation}
\frac{\delta \mathcal{E}}{\delta a} = - G[a(\theta)] + G_\mathrm{c}(\theta,a(\theta)).
\end{equation}
The crack front dynamics can therefore be recast as the gradient flow
\begin{equation}
\frac{\partial a}{\partial t} \propto - \frac{\delta \mathcal{E}}{\delta a}.
\label{eq_motion}
\end{equation}

Expanding the total energy around stationary configuration $a_\mathrm{s}$ yields
\begin{equation}
\mathcal{E}[a] = \mathcal{E}[a_\mathrm{s}] + \frac12
 \left.\frac{\delta^2 \mathcal{E}}{\delta a^2}\right|_{a_\mathrm{s}}
 (a - a_\mathrm{s})^2
\end{equation}
where the linear term vanishes because Griffith-compatible stationary configurations are energy minima.

Using this expansion into Eq.~\eqref{eq_motion} gives
\begin{equation}
\frac{\partial a}{\partial t} \propto \left. \frac{\delta^2 \mathcal{E}}{\delta a^2} \right|_{a_\mathrm{s}} (a_\mathrm{s} - a)
\label{eq_motion2}
\end{equation}
showing that the relaxation rate is controlled by the curvature of the energy landscape at the stationary state.

We then perturb the crack front according to
\begin{equation}
a(\theta,t) = a_\mathrm{s}(\theta) + \delta a(\theta,t)
\end{equation}
assuming $\delta a(\theta,0)  \ll a_\mathrm{s}(\theta)$. Equation~\eqref{eq_motion2} then becomes
\begin{equation}
\frac{\partial \delta a}{\partial t} \propto - \left. \frac{\delta^2 \mathcal{E}}{\delta a^2} \right|_{a_\mathrm{s}} \delta a
\label{eq_motion3}
\end{equation}
whose solution is the exponential relaxation
\begin{equation}
\delta a(\theta,t) \propto e^{-t/\tau}
\end{equation}
with relaxation time
\begin{equation}
\tau \propto  \left (\left. \frac{\delta^2 \mathcal{E}}{\delta a^2} \right|_{a_\mathrm{s}} \right)^{-1}.
\label{time_relax}
\end{equation}

The relaxation time is controlled by the curvature of the energy minimum shown in the energy landscapes of Fig.~3 of the main manuscript. As the fingering threshold is approached ($\Delta \rightarrow \Delta_\mathrm{c}$), the energy well progressively flattens and ultimately disappears at the bifurcation point, causing the relaxation time to diverge. Expanding the energy curvature about the critical crack-front configuration $a_\mathrm{c}$ yields
\begin{equation}
\begin{split}
\left. \frac{\delta^2 \mathcal{E}}{\delta a^2}\right |_{a_\mathrm{s}} & = \left. \frac{\delta^2 \mathcal{E}}{\delta a^2}\right |_{a_\mathrm{c}} + \left. \frac{\delta^3 \mathcal{E}}{\delta a^3}\right |_{a_\mathrm{c}} (a_\mathrm{s} - a_\mathrm{c}) \\
& = \left. \frac{\delta^3 \mathcal{E}}{\delta a^3}\right |_{a_\mathrm{c}} (a_\mathrm{s} - a_\mathrm{c})
\end{split}
\end{equation}
where the second derivative vanishes at the bifurcation point. Because the curvature decreases as the stationary configuration approaches $a_\mathrm{c}$, the third derivative at the bifurcation point is negative. Consequently,
\begin{equation}
\left. \frac{\delta^2 \mathcal{E}}{\delta a^2}\right |_{a_\mathrm{s}} \propto (a_\mathrm{c} - a_\mathrm{s}),
\end{equation}
and Eq.~\eqref{time_relax} immediately leads
\begin{equation}
\tau \propto 1/(a_\mathrm{c} - a_\mathrm{s}).
\end{equation}

Using the square-root scaling
\begin{equation}
a_\mathrm{c} - a_\mathrm{s} \propto \sqrt{\Delta_\mathrm{c} - \Delta}
\end{equation}
demonstrated in the previous section and observed numerically in Fig.~4(b) of the main manuscript finally yields
\begin{equation}
\tau \propto  1/\sqrt{\Delta_\mathrm{c} - \Delta}.
\end{equation}

The resulting square-root critical slowing down reproduces one of the hallmark signatures of classical saddle-nod bifurcations, yet arising from a fundamentally different mechanism: the disappearance of Griffith-compatible crack-front equilibria rather than the annihilation of stable and unstable branches.

\section{Griffith-compatible stable branch as the unique attractor}
\begin{figure}[h]
\centerline{\includegraphics[width=1.1\linewidth]{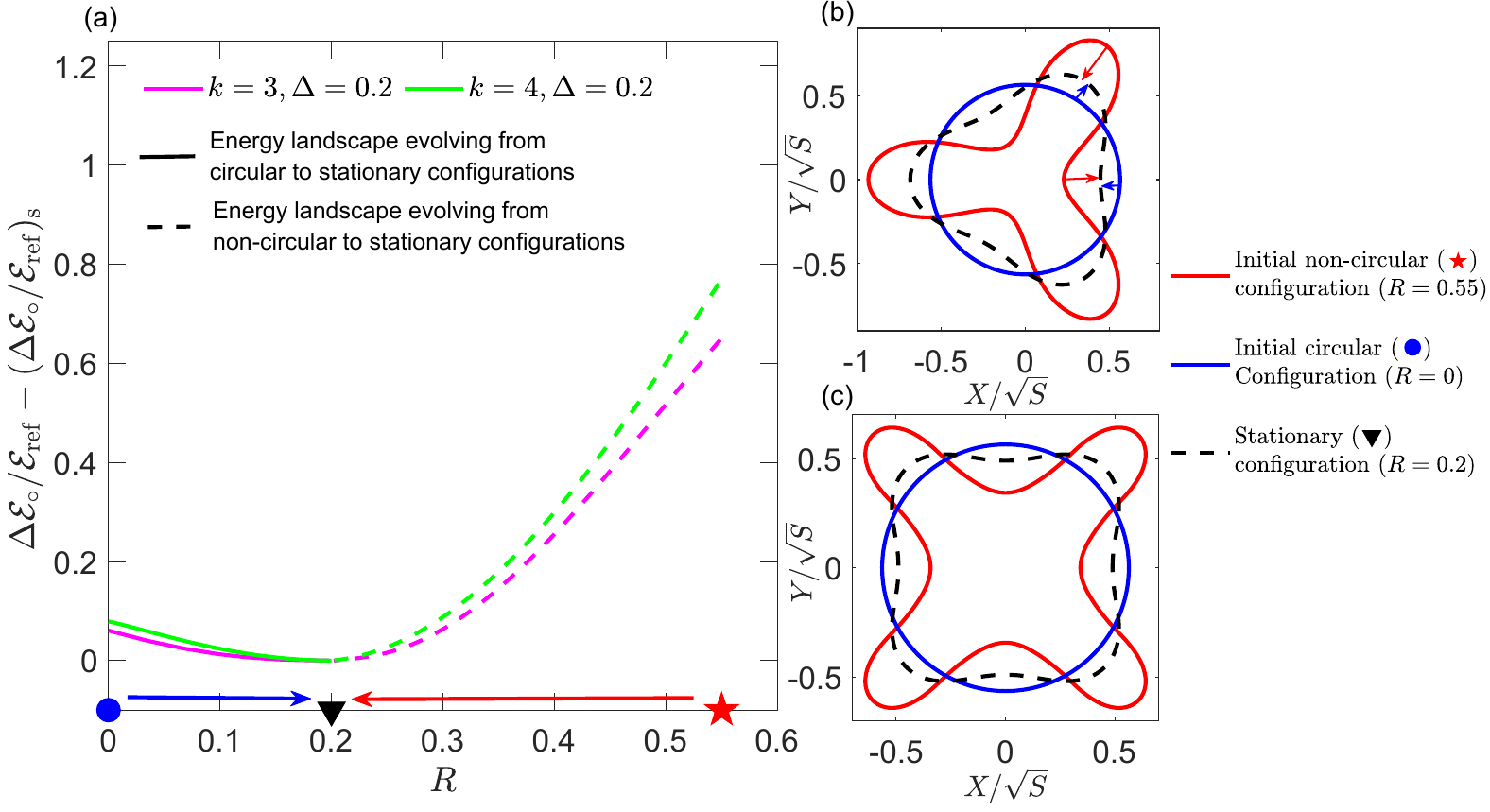}}
\caption{Demonstration of the stationary configuration (Griffith-compatible stable branch) as the unique attractor by showing energy landscapes on panel (a) for $k=3 \ {\rm and}\ 4$ with $\Delta=0.2$: {\it solid} lines represent energy landscapes evolving from the circular crack configuration ($\tikzsymbol{fill=blue}$) to the stationary configuration (\triad{}) and {\it dashed} lines are for energy landscape evolves from the non-circular crack configuration ({\color{red}$\bigstar$}) to the stationary configuration (shown by {\it red} arrow). Panels (b) and (c) plot these three crack configurations for $k=3$ and $k=4$, respectively: Initial non-circular crack configuration ($R=0.55$), Initial circular crack configuration ($R=0$), and stationary crack configuration ($R=0.2$). Irrespective of the initial crack configuration, it always evolves to the stationary configuration for which Griffith's condition is satisfied at all points along the crack front.}
\label{fig:app4}
\end{figure}

This appendix demonstrates that the Griffith-compatible stable branch corresponding to stationary crack configurations is the unique attractor. We consider two initial crack configurations (as shown in Fig. \ref{fig:app4}b and c) and deform them using a sinusoidal toughness field, $K_c = \overline{K}_c [1+\Delta\cos(k\theta)]$ with $k=3,4,\ {\rm and}\ \Delta=0.2$. Fig. \ref{fig:app4}a plots the corresponding energy landscapes for these two cases: circular crack configuration as starting point (solid lines) and non-circular ($k-$ petal shape) crack configuration (dashed lines). In both cases, which are far away in their geometry (one with $R=0$ and the other with $R=0.55$), the energy landscape evolves to the same stationary crack configuration ($R=0.2$) in which Griffith's condition is satisfied along the entire crack front. Therefore, this shows that the Griffith-compatible stable branch is the unique attractor. 

\section{Universal geometry of the crack front at the fingering threshold}

This appendix establishes the universal character of the crack-front geometries shown in Fig. 4d of the main text. To reveal this universality, we rescale both the radial and angular coordinates. Specifically, we subtract the minimum radius, $a(\theta)$, from the crack-front profile $a(0)$ and normalize he resulting radial displacement by the perturbation amplitude, $\Delta a$. The rescaled radial coordinate $[a(0)-a(0)]/\Delta a$ is plotted against the angular coordinate normalized by the perturbation period, $\theta/(2\pi/k)$.
\begin{figure}[h!]
\centerline{\includegraphics[width=1\linewidth]{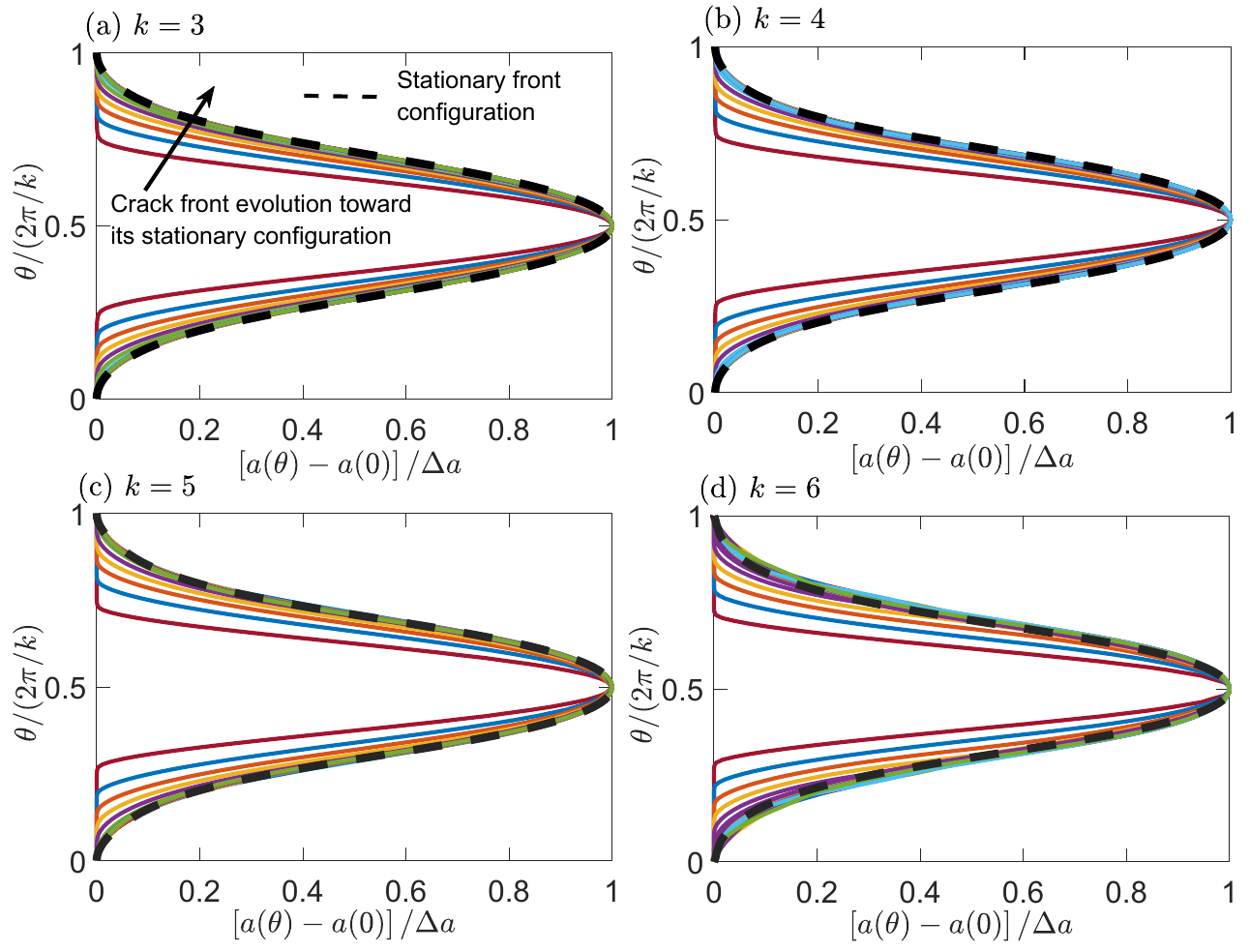}}
\caption{Evolution of the normalized front geometry $[a(\theta)-a(0)]/\Delta a$ during propagation at the fingering threshold, $\Delta = \Delta_\mathrm{c}$, for toughness maps  containing (a) $k=3$, (b) $k=4$, (c) $k=5$, and (d) $k=6$ obstacles. The four stationary profiles collapse onto a single master curve shown in Fig.~4d of the main text, demonstrating that the normalized crack-front geometry is independent of $k$ at threshold.}
\label{fig:app5}
\end{figure}

As shown in Fig.~\ref{fig:app5}, the normalized crack front initially deforms through a transient regime, as discussed in the main text, before reaching a stationary configuration (dashed black curves). Remarkably, the stationary profiles obtained for different values of $k$ collapse onto a single curve, revealing a universal stationary crack-front shape independent of $k$. The resulting master curve is reported in Fig.~4d of the main text.
\bibliography{bibfracture}